# Unipolar and bipolar fatigue in antiferroelectric lead zirconate thin films and evidences for switching-induced charge injection inducing fatigue

X.J. Lou[*] and J. Wang

*Department of Materials Science and Engineering, National University of Singapore, Singapore 117574, Singapore*

**Abstract:**

For the first time, we show that unipolar fatigue does occur in antiferroelectric capacitors, confirming the predictions of a previous work [Appl. Phys. Lett., **94**, 072901 (2009)]. We also show that unipolar fatigue in antiferroelectrics is less severe than bipolar fatigue if the driving field is of the same magnitude. This phenomenon has been attributed to the switching-induced charge injection, the main cause for polarization fatigue in ferroelectric and antiferroelectric materials. Other evidences for polarization fatigue caused by the switching-induced charge injection from the nearby electrode rather than the charge injection during stable/quasi-stable leakage current stage are also discussed.

---

[*] Author to whom correspondence should be addressed; electronic mail: mselx@nus.edu.sg

Recently, antiferroelectric materials have attracted much research interest due to their potential applications in microactuators and energy conversion devices.[1] Antiferroelectrics show characteristic double hysteresis loop upon the application of an ac external field, different from the well known single loop shown by ferroelectrics. However, like ferroelectric materials antiferroelectrics also show polarization fatigue under repetitive bipolar electrical cycling, which has been one of the main obstacles for the development of electronic devices made from these materials.

For many years, polarization fatigue also represents one of the most serious difficulties for realizing nonvolatile ferroelectric random access memories (FeRAMs),[2,3] along with imprint,[2] polarization retention[4,5] and electrical breakdown.[6,7] Although polarization fatigue in ferroelectric materials has been extensively studied during the last few decades, that in antiferroelectrics has not. Furthermore, the very limited publications on fatigue in antiferroelectric capacitors are all concentrated on bipolar fatigue and surprisingly enough there is *no* work on unipolar fatigue in antiferroelectric materials.

Recently, we have built a fatigue theory for ferroelectric materials, which emphasizes the extremely high depolarization field generated near the electrode-film interface by the polarization charges at the tip of the needlelike domains during switching (therefore we called it the LPD-SICI model; LPD-SICI denotes local phase decomposition caused by switching-induced charge injection).[3,8,9] This model has been subsequently generalized to describe the fatigue problem in antiferroelectrics.[10] In particular, this model explains why polarization fatigue in antiferroelectric materials is much less severe than that in ferroelectrics under similar or even more severe bipolar electrical cycling as observed in the literature.[10-13] Moreover, this model indicates that antiferroelectric materials should show observable fatigue under unipolar electrical cycling, which has not been confirmed yet so far due to the lack of experimental data in the literature as mentioned above. In this work, we show unambiguously that both bipolar fatigue and unipolar fatigue occur in antiferroelectric lead zirconate (PZO) thin films with the latter being less severe than the former under cycling field of the same magnitude. The implications of the present work and the evidences for switching-induced charge injection causing fatigue are also discussed.

The PZO thin films (~370 nm in thickness) used in the present work were fabricated on Pt/TiO$_x$/SiO$_2$/Si substrates using sol-gel spin coating method. 10% Pb-excess solutions of 0.4 M in concentration was used to prepare the films in order to compensate Pb loss during annealing. After a certain number of layers were spin-coated, the PZO films were finally annealed at 650 ºC, 675 ºC and 700 ºC, respectively, for 15 min in air in a quartz tube furnace (Carbolyte) to achieve the crystallized phase. The one annealed at 675 ºC for 15 min shows the best antiferroelectric properties in comparison with the other two and was therefore used in the present work. X-Ray Diffraction (XRD, D8 Advanced Diffractometer System, Bruker) and Field Emission Scanning Electron (FE-SEM, XL30 FEG Philips) studies show that the films are polycrystalline and of a pure perovskite phase without secondary phases. For electrical measurements, the top electrodes (Pt squares of ~100 x 100 μm$^2$) were then deposited by dc sputtering on the films via transmission electron microscopy (TEM) grids. Polarization fatigue measurements (i.e., the standard cycling-and-PUND procedure) and hysteresis-loop measurements (at 1 kHz) were conducted using a Radiant ferroelectric test system. Bipolar/unipolar triangle wave of $10^5$ Hz was used for the fatigue tests. Dielectric measurements were carried out using an impedance analyzer (Solartron SI 1260).

Fig 1 (a-i) show the bipolar and unipolar fatigue properties of our antiferroelectric PZO thin film. Bipolar fatigue data at ±17 V evaluated by hysteresis-loop measurements are shown in Fig 1 (a), and those evaluated by dielectric measurements are shown in Fig 1 (b)(c). Fig 1 (b) shows the dielectric constant and tangent loss as a function of measurement frequency for 1, 4x10$^5$, 1.9x10$^6$, 1.15x10$^7$, 1.02x10$^8$ and 1x10$^9$ cycles. Fig 1 (c) displays the dielectric constant measured at 10$^2$ and 10$^5$ Hz as a function of cycle number $N$. Fig 1 (d-f) and Fig 1 (g-i) show unipolar fatigue data on another two virgin capacitors in the same film at -17 V and +17 V, respectively, evaluated by the same methods as those shown in Fig 1 (a-c). Detailed descriptions of the curves are given in the captions of Fig 1 (note that the 10$^9$-cycle data for unipolar fatigue at -17 V were not obtained, because all the tested virgin capacitors brokedown before 10$^9$ cycles were reached). In order to check the repeatability of the data shown in Fig 1, we repeated each measurement on another 3-5 virgin capacitors in the same film, which shows that the difference between the data collected from one measurement and those from another is negligible, confirming that the profiles shown in Fig 1 (a-i) indeed represent the real characters of bipolar and/or unipolar fatigue of our antiferroelectric PZO thin-film capacitors. We noticed that the hysteresis loops of the virgin capacitors are a little asymmetric [see Fig 1 (a)(d)(g)], which is probably because of the different thermal histories experienced by the top and bottom electrodes and consequently the different interface and phase-nucleation/switching properties. Asymmetric hysteresis loops are indeed common for antiferroelectric PZO thin films and were also reported by Kim et al.[14] and Jang et al.[15] The slight asymmetric loop does not affect the conclusions of this work as we will see shortly, because we did both of the two possible unipolar fatigue measurements: along one branch of the loop at -17 V, and along the other at +17 V [see Fig 1 (d-f) and Fig 1 (g-i)].

Fig 1(a) shows that the hysteresis loop loses its squareness to a certain degree upon progressive bipolar fatigue at ±17 V. The "coercive fields" for both of the branches (defined in a similar way as that for ferroelectrics' single loop) increase significantly with $N$, indicating that fatigue occurs in the PZO film during repetitive bipolar cycling. The bipolar fatigue properties of our PZO film could also be seen from the dielectric data: Fig 1 (b) shows that the higher the cycle number $N$ the lower the profile of dielectric constant versus frequency will be; Fig 1 (c) shows that the dielectric constant measured at $10^5$ Hz (or $10^2$ Hz) decreases from ~120 for $N=1$ to ~97 for $N=10^9$, that is, ~19% of the original dielectric constant was lost after $10^9$ bipolar fatigue cycles, which is still less severe than the dielectric constant loss reported for the ferroelectric materials undergoing bipolar fatigue.[16-19] Therefore, the dielectric data shown here and previously[16-19] are in good agreement with the polarization or hysteresis-loop results:[11-13] both show that antiferroelectrics show higher fatigue resistance than ferroelectric materials under bipolar electrical cycling, consistent with the predictions of the LPD-SICI model.[10]

Fig 1 (d-f) and (g-i) show, for the first time to our best knowledge, that electrical fatigue indeed occurs for antiferroelectrics under unipolar cycling, confirming the predictions of our previous work.[10] One can also see that both of the branches of the loop shrink slightly under unipolar fatigue [Fig 1 (d) and (g)], independent of polarity (i.e., -17 V or +17 V). We believe that it is mainly due to the polarity-independent field-screening effect of the degraded layer(s): that is, the field seen by the bulk film is reduced after interfacial decomposition because of the low dielectric constant of the degraded layer.

Comparing Fig 1 (d)(g) with Fig 1 (a), Fig 1 (e)(h) with Fig 1(b) and Fig 1 (f)(i) with Fig 1(c), we see that unipolar fatigue in antiferroelectrics is less severe than bipolar fatigue driven by electric fields of the same magnitude. The same phenomenon has also been reported for ferroelectric materials.[3,20-22] The less severity of unipolar fatigue in comparison with bipolar fatigue in antiferroelectrics can also be clearly seen from Fig 2 (a)(b): both the less significant increase in $P_r(N)/P_r(0)$ as a function of $N$ in Fig 2 (a) and the less noteworthy decrease in the normalized dielectric constant as a function of $N$ in Fig 2 (b) during unipolar fatigue at -17 V or +17 V in comparison with those during bipolar fatigue at ±17 V suggest that antiferroelectric materials show higher resistance to unipolar fatigue than to bipolar one, provided that the driving field is of the same magnitude.

The better unipolar fatigue endurance compared with bipolar one is attributed to the switching-induced charge injection from the nearer electrode, which is the main cause for local phase decomposition and consequently polarization fatigue in both antiferroelectrics and ferroelectrics. That is because we should observe similar fatigue resistances appearing for both bipolar and unipolar fatigue if they are caused by the charge injection during the stable/quasi-stable leakage current stage. But that is not the case.

Other evidences for polarization fatigue caused by the switching-induced charge injection rather than the charge injection during leakage stage include the following experimental observations: (1) the optical/thermal fatigue data of Warren *et al.*[23] [see, particularly, Fig 3 (a) in Ref [23] and the discussion in Ref [9]]; (2) the studies of the effect of a dc field just below coercive field $E_c$ before polarization reversal by Colla *et al*. [see, particularly, Fig 3 in Ref [24] and the discussion in Ref [9]]; (3) the fact that unipolar pulses produce no or less severe fatigue than bipolar pulses in ferroelectrics [see Ref [3,9,19]]; (4) the fact that antiferroelectrics show better fatigue endurance than ferroelectrics under bipolar electrical cycling [see Ref [10]]; (5) the well-known phenomena that slightly leaky lead zirconate titanate (PZT) capacitors electroded with conductive oxides (e.g. $RuO_2/PZT/RuO_2$) show improved fatigue endurance (or even fatigue free) than those with metal electrodes (e.g. Pt/PZT/Pt) that are more insulating;[2] (6) the orientation-dependent (and therefore polarization-dependent) fatigue studies of Takemura *et al* on ferroelectric single crystals [see, particularly, Fig 12 in Ref [25] and the discussion in Ref [9]].

In summary, we show, for the first time, that unipolar fatigue occurs in antiferroelectrics, confirming the predictions of a previous work.[10] Furthermore, it has been shown that unipolar fatigue in antiferroelectric capacitors is less severe than bipolar fatigue provided that the magnitude of the cycling field is about the same. We attributed this phenomenon to the switching-induced charge injection, which is the main reason for electrical fatigue in both ferroelectric and antiferroelectric materials. Finally, we discussed several experimental observations or evidences that support a scenario that polarization fatigue is mainly caused by the switching-induced charge injection rather than the charge injection during stable/quasi-stable leakage stage.

X.J.L would like to thank the LKY PDF established under the Lee Kuan Yew Endowment Fund for support. The work is supported also by National University of Singapore and a MOE AcRF grant (R284-000-058-112).

**Figure Captions:**

Fig 1 (color online) Bipolar and unipolar fatigue properties of our antiferroelectric PZO thin film. Bipolar fatigue at ±17 V was evaluated by (a) hysteresis-loop measurements and (b)(c) dielectric measurements. (b) shows dielectric constant and tangent loss as a function of measurement frequency for 1, $4 \times 10^5$, $1.9 \times 10^6$, $1.15 \times 10^7$, $1.02 \times 10^8$ and $1 \times 10^9$ cycles; (c) displays dielectric constant measured at $10^2$ and $10^5$ Hz as a function of cycle number $N$. Similarly, unipolar fatigue data at -17 V evaluated by (d) hysteresis-loop and (e)(f) dielectric measurements, and unipolar fatigue data at +17 V evaluated by (g) hysteresis-loop and (h)(i) dielectric measurements are also shown.

Fig 2 (color online) Comparison of unipolar fatigue and bipolar fatigue in the antiferroelectric PZO thin film: (a) $P_s(N)/P_s(0)$ and $P_r(N)/P_r(0)$ as a function of $N$, and (b) the normalized dielectric constant measured at $10^5$ Hz as a function of $N$.

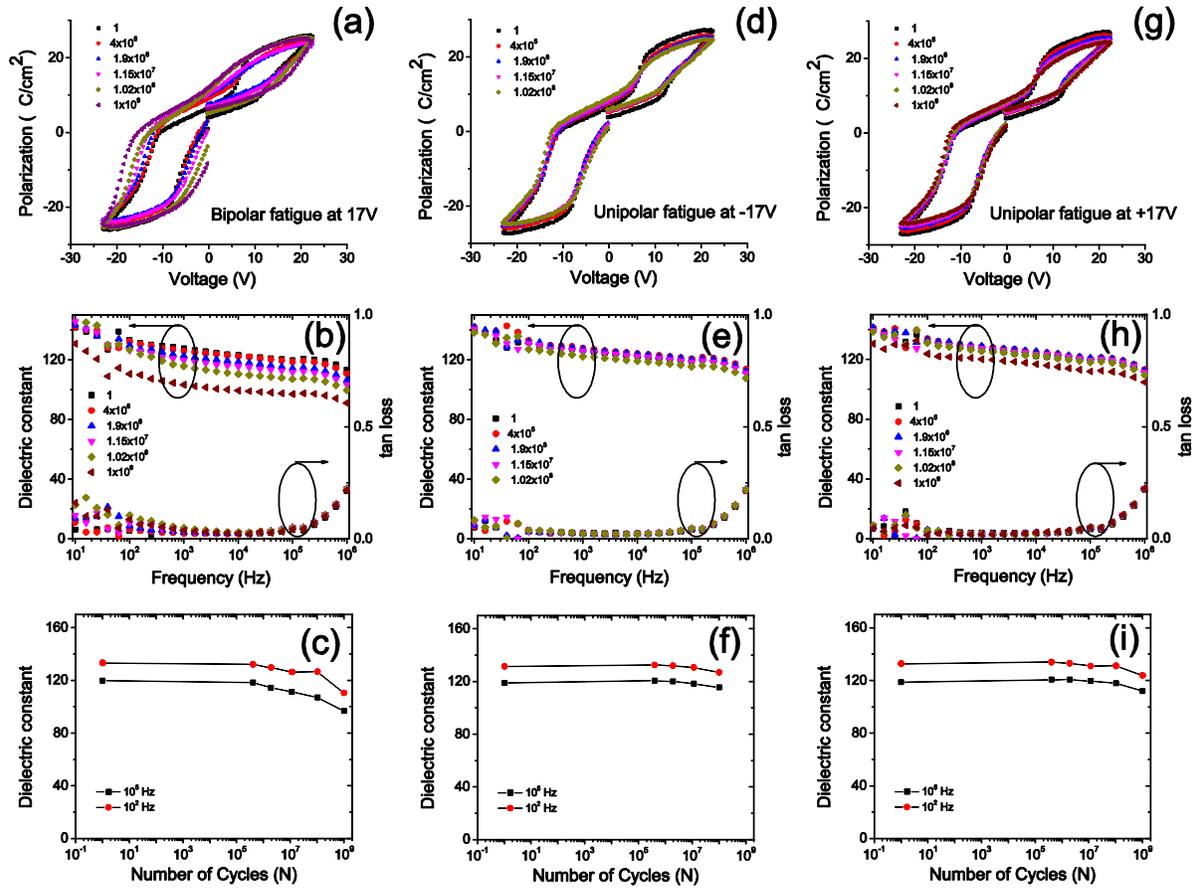

Fig 1

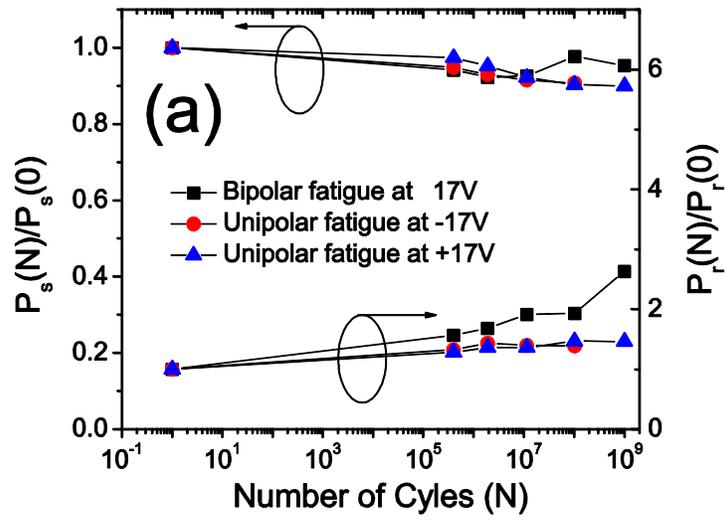

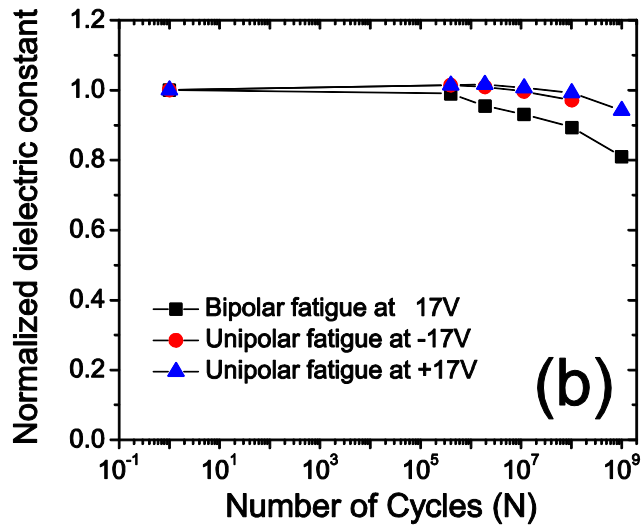

Fig 2